\documentclass[12pt,draftcls,journal,onecolumn]{IEEEtran}
\usepackage{graphicx}
\usepackage{epstopdf}
\usepackage[latin1]{inputenc}
\usepackage{amsmath,amssymb,amscd,latexsym,dsfont}
\usepackage[english]{babel}
\usepackage{tabularx,cite}
\usepackage{graphicx}
\usepackage{algpseudocode}
\usepackage{algorithm}
\usepackage{algorithmicx}
\usepackage{float}
\usepackage{multicol}
\usepackage{psfrag}
\usepackage{comment,psfrag,subfigure,enumerate}
\usepackage{color}
\usepackage{amsthm}
\ifCLASSINFOpdf
\else
\fi
\begin{document}
%
\title{On the Required Number of Antennas in a Point-to-Point Large-but-Finite MIMO System: Outage-Limited Scenario}
\author{\IEEEauthorblockN{Behrooz Makki, Tommy Svensson, Thomas Eriksson and Mohamed-Slim Alouini, \emph{Fellow, IEEE}}\\
\thanks{Behrooz Makki, Tommy Svensson and Thomas Eriksson are with Chalmers University of Technology, Email: \{behrooz.makki, tommy.svensson, thomase\}@chalmers.se. Mohamed-Slim Alouini is with the King Abdullah University of Science and Technology (KAUST), Email: slim.alouini@kaust.edu.sa}
\thanks{Part of this work has been submitted for possible presentation at the IEEE ICUWB 2015.}
}

%
\maketitle
\vspace{-16mm}
\begin{abstract}
This paper investigates the performance of the point-to-point multiple-input-multiple-output (MIMO) systems in the presence of a large but finite numbers of antennas at the transmitters and/or receivers. Considering the cases with and without hybrid automatic repeat request (HARQ) feedback, we determine the minimum numbers of the transmit/receive antennas which are required to satisfy different outage probability constraints. Our results are obtained for different fading conditions and the effect of the power amplifiers efficiency on the performance of the MIMO-HARQ systems is analyzed. Moreover, we derive closed-form expressions for the asymptotic performance of the MIMO-HARQ systems when the number of antennas increases. Our analytical and numerical results show that different outage requirements can be satisfied with relatively few transmit/receive antennas.
\end{abstract}
%
\IEEEpeerreviewmaketitle
\vspace{-6mm}
\section{Introduction}
The next generation of wireless networks must provide data streams for everyone
everywhere at any time. Particularly, the data rates should be orders of magnitude higher than those in the current systems; a demand that creates serious power concerns because the data rate scales with power monotonically. The problem becomes even more important when
we remember that currently the wireless network contributes $\sim2\%$ of global $\text{CO}_2$ emissions and its energy consumption is expected to increase $16-20\%$ every year \cite{Gartner}.

To address the demands, the main strategy persuaded in the last few years is the network \emph{densification} \cite{6736747}. One of the promising techniques to densify the network is to use many antennas at the transmit and/or receive terminals. This approach is referred to as massive or large multiple-input-multiple-output (MIMO) in the literature.

In general, the more antennas the transmitter and/or the receiver are equipped with,
the better the data rate/link reliability. Particularly, the capacity increases and the required uplink/downlink transmit power decreases with the number of antennas.
Thus, the trend is towards  asymptotically high number of antennas. This is specially because  millimeter wave communication \cite{6515173,5783993}, which are indeed expected to be implemented in the next generation of wireless networks, makes it possible to assemble many antennas at the transmit/receive terminals. However, large MIMO implies challenges such as hardware impairments and signal processing complexity which may limit the number of antennas in practice. Also, one of the main bottlenecks of large MIMO is the channel state information (CSI) acquisition, specifically at the transmitter. Therefore, it is interesting to use efficient feedback schemes such as hybrid automatic repeat request (HARQ) whose feedback overhead does not scale with the number of antennas.


The performance of HARQ protocols in single-input-single-output (SISO) and MIMO systems is studied in, e.g., \cite{Tcomkhodemun,throughputdef,outageHARQ,6487360} and \cite{a01661837,a04608960,a05351683,a06035974,6918481,MIMOARQkhodemun,4411583,4601023,4663876,6006606,a05513476,b06120361}, respectively. MIMO transmission with many antennas is advocated in \cite{4176578,5595728}
where the time-division duplex (TDD)-based training is utilized for CSI feedback\footnote{The results of \cite{4176578,5595728,5947131,5205814,5502810,5898372,6824794,6415388,5990435,6777295,6816089,6678040,6120314,6241389,6172680,6736761,6798744,6375940} are mostly on multi-user MIMO networks, as opposed to our work on point-to-point systems. However, because many of the analytical results in \cite{4176578,5595728,5947131,5205814,5502810,5898372,6824794,6415388,5990435,6777295,6816089,6678040,6120314,6241389,6172680,6736761,6798744,6375940}  are applicable in point-to-point systems as well, these works are cited.}.
Also,\cite{5947131,5205814,5502810,5898372,6824794,6415388,5990435} introduce TDD-based schemes for large systems.
In the meantime, frequency division duplex (FDD)-based massive MIMO has recently attracted attentions and low-overhead CSI acquisition methods were proposed \cite{6777295,6816089,6678040}. Considering imperfect CSI, \cite{6120314} derives lower bounds for the uplink achievable rate of the MIMO setups with large but finite number of antennas. Finally, \cite{6241389} (resp. \cite{6172680}) studies the zero-forcing based TDD (resp. TDD/FDD) systems under the assumption that the number of transmit antennas and the single-antenna users are asymptotically large while their ratio remains bounded (For detailed review of the literature on massive MIMO, see \cite{6736761,6798744,6375940}).

To summarize, a large part of the literature on the point-to-point and multi-user large MIMO is based on the assumption of asymptotically many antennas. Then, a natural question is
how many transmit/receive antennas do we require in practice to satisfy different quality-of-service requirements. The interesting answer this paper establishes is relatively few, for a large range of outage probabilities.

Here, we study the outage-limited performance of point-to-point MIMO systems in the cases with large but finite number of antennas. We derive closed-form expressions for the required number of transmit and/or receive antennas satisfying various outage probability requirements (Theorem 1). The results are obtained for different fading conditions and in the cases with or without HARQ. Furthermore, we analyze the effect of the power amplifiers (PAs) efficiency and the antennas spatial correlation on the system performance (Sections IV.A and V.B, respectively) and study the outage probability in the cases with adaptive power allocation between the HARQ retransmissions (Section IV.B). Finally, we study the asymptotic performance of MIMO systems. Particularly, denoting the outage probability and the number of transmit and receive antennas by $\Pr(\text{Outage}), N_\text{t}$ and $N_\text{r,}$ respectively, we derive closed-form expressions for the normalized outage factor which is defined as $\Gamma=-\frac{\log(\Pr(\text{Outage}))}{N_\text{t}N_\text{r}}$ when the number of transmit and/or receive antennas increases (Theorem 2).


As opposed to \cite{Tcomkhodemun,throughputdef,outageHARQ,6487360,a01661837,a04608960,a05351683,a06035974,6918481,MIMOARQkhodemun,4411583,4601023,4663876,6006606,a05513476,b06120361}, we consider large MIMO setups and
determine the required number of antennas in outage-limited conditions. Also, the paper is different from \cite{4176578,5595728,5947131,5205814,5502810,5898372,6824794,5990435,6415388,6777295,6816089,6678040,6120314,6172680,6241389,6736761,6798744,6375940} because we study the outage-limited scenarios in point-to-point systems, implement HARQ and the number of antennas is considered to be finite. The differences in the problem formulation and the channel model makes the problem solved in this paper completely different from the ones in \cite{Tcomkhodemun,throughputdef,outageHARQ,6487360,a01661837,a04608960,a05351683,a06035974,6918481,MIMOARQkhodemun,4411583,4601023,4663876,6006606,a05513476,b06120361,4176578,5595728,5947131,5205814,5502810,5898372,6824794,5990435,6415388,6777295,6816089,6678040,6120314,6172680,6241389,6736761,6798744,6375940}, leading to different analytical/numerical results, as well as to different conclusions. Finally, our discussions on the asymptotic outage performance of the MIMO setups and the effect of PAs on the performance of MIMO-HARQ schemes have not been presented before.

Our analytical and numerical results indicate that:
\begin{itemize}
  \item Different quality-of-service requirements can be satisfied with relatively few transmit/receive antennas. For instance, consider a SIMO (S: single) setup without HARQ and transmission signal-to-noise ratio (SNR) $5$ dB. Then, with the data rate of 3 nats-per-channel use (npcu), the outage probabilities $10^{-3}$, $10^{-4}$ and $10^{-5}$ are guaranteed with $16,$ $18$ and $20$ receive antennas, respectively (Fig. 5a). Also, the implementation of HARQ reduces the required number of antennas significantly (Fig. 3).
  \item Considering the moderate/high SNRs, the required number of transmit (resp. receive) antennas scales with $(Q^{-1}(\theta))^2,$ $\frac{1}{MT},$ $\frac{1}{N_\text{r}},$ and $\frac{1}{(\log(\phi))^2}$ linearly, if the number of receive (resp. transmit) antennas is fixed. Here, $M$ is the maximum number of HARQ retransmission rounds, $T$ is the number of channel realizations experienced in each round, $\phi$ denotes the SNR, $\theta$ is the outage probability constraint and $Q^{-1}(.)$ represents the inverse Gaussian $Q$-function. These scaling laws are changed drastically, if the numbers of the transmit and receive antennas are adapted simultaneously (see Theorem 1 and its following discussions for details).
  \item For different fading conditions, the normalized outage factor $\Gamma=-\frac{\log(\Pr(\text{Outage}))}{N_\text{t}N_\text{r}}$ converges to constant values, unless the number of receive antennas grows large while the number of transmit antennas is fixed (see Theorem 2). Also, for every given number of transmit/receive antennas, the normalized outage factor increases with $MT$ linearly.
  \item There are mappings between the performance of MIMO-HARQ systems in quasi-static, slow- and fast-fading conditions, in the sense that with proper scaling of the channel parameters the same outage probability is achieved in these conditions. This point provides an appropriate connection between the papers considering one of these fading models.
  \item Adaptive power allocation between the HARQ retransmissions leads to marginal antenna requirement reduction, while the system performance is remarkably affected by the inefficiency of the PAs. Finally, the spatial correlation between the antennas increases the required number of antennas while, for a large range of correlation conditions, the same scaling rules hold for the uncorrelated and correlated fading scenarios.
\end{itemize}

\vspace{-0mm}
 \vspace{-0mm}
\section{System Model}
Consider a point-to-point MIMO setup with $N_\text{t}$ transmit antennas and $N_\text{r}$ receive antennas. We study the block-fading conditions where the channel coefficients remain constant during the channel coherence time and then change to other values based on their probability density function (PDF). In this way, the received signal is given by
\begin{align}\label{eq:newchannelmodel}
{\textbf{Y}} = \textbf{H}{\textbf{X}} + {\textbf{Z}},{\textbf{Z}} \in \mathcal{C}^{N_\text{r}\times 1},
\end{align}
where $\textbf{H}\in \mathcal{C}^{N_\text{r}\times N_\text{t}}$ is the fading matrix, $\textbf{X}\in\mathcal{C}^{N_\text{t}\times 1}$ is the transmitted signal and $\textbf{Z}\in \mathcal{C}^{N_\text{r}\times 1}$ denotes the independent and identically distributed (IID) complex Gaussian noise matrix. The results are mainly given for IID Rayleigh-fading channels where each element of the channel matrix $\textbf{H}$ follows a complex Gaussian distribution $\mathcal{CN}(0,1)$ (To analyze the effect of the antennas spatial correlation, see Fig. 7 and Section V.B).
The channel coefficients are assumed to be known at the receiver which is an acceptable assumption in block-fading channels \cite{Tcomkhodemun,throughputdef,outageHARQ,6487360,a01661837,a04608960,a05351683,a06035974,6918481,MIMOARQkhodemun,4411583,4601023,4663876,6006606,a05513476,b06120361}. On the other hand, there is no CSI available at the transmitter except the HARQ feedback bits.
The feedback channel is supposed to be delay- and error-free.

As the most promising HARQ approach leading to highest throughput/lowest outage probability \cite{Tcomkhodemun,throughputdef,MIMOARQkhodemun,a01661837}, we consider the incremental redundancy (INR) HARQ with a maximum of $M$ retransmissions, i.e., the message is retransmitted a maximum of $M$ times. Note that setting $M=1$ represents the cases without HARQ, i.e., open-loop communication. Also, a packet is defined as the transmission of a codeword along with all its possible retransmissions. We investigate the system performance for three different fading conditions:
\begin{itemize}
  \item \textbf{Fast-fading}. Here, it is assumed that a finite number of channel realizations are experienced within each HARQ retransmission round.
  \item \textbf{Slow-fading}. In this model, the channel is supposed to change between two successive retransmission rounds, while it is fixed for the duration of each retransmission.
  \item \textbf{Quasi-static}. The channel is assumed to remain fixed within a packet period.
\end{itemize}
Fast-fading is an appropriate model for fast-moving users or users with long codewords compared to the channel coherence time \cite{6510028,MIMOARQkhodemun}. On the other hand, slow-fading can properly model the cases with users of moderate speeds or frequency-hopping schemes \cite{a04608960,a05351683,a06035974,4411583,4601023,4663876}. Finally, the quasi-static represents the scenarios with slow-moving or stationary users, e.g., \cite{Tcomkhodemun,6006606,a05513476,b06120361,a01661837,4663876}.


\section{Problem Formulation}
Considering the INR HARQ with a maximum of $M$ retransmissions, $Q$ information nats are encoded into a \emph{parent} codeword of length $ML$ channel uses and the codeword is divided into $M$ sub-codewords of length $L$. In each retransmission round, the transmitter sends a new sub-codeword and the receiver combines all signals received up to the end of that round. Thus, the equivalent rate at the end of round $m$ is $\frac{Q}{mL}=\frac{R}{m}$ npcu where $R$ denotes the initial transmission rate. The retransmissions continue until the message is correctly decoded by the receiver or the maximum permitted retransmission round is reached.

Let us denote the determinant and the Hermitian of the matrix $\textbf{X}$ by $|\textbf{X}|$ and $\textbf{X}^\text{h}$, respectively. Assuming fast-fading conditions with $T$ independent fading realizations $\textbf{H}((m-1)T+1), \ldots, \textbf{H}(mT)$ in the $m$th round and an isotropic Gaussian input distribution over all transmit antennas, the results of, e.g., \cite[Chapter 15]{4444444444}, \cite[Chapter 7]{ELGAMAL}, can be used to find the outage probability of the INR-based MIMO-HARQ scheme as
\begin{align}\label{eq:outageprobinrfast}
\Pr(\text{Outage})^\text{Fast-fading}=\Pr\left(\frac{1}{MT}\sum_{n=1}^{M}\sum_{t=(n-1)T+1}^{nT}\log\left|\textbf{I}_{N_\text{r}}+\frac{\phi}{N_\text{t}}\textbf{H}(t)\textbf{H}(t)^h\right|\le \frac{R}{M}\right).
\end{align}
Here, $\phi$ is the total transmission power and $\frac{\phi}{N_\text{t}}$ is the transmission power per transmit antenna (in dB, we have $10\log_{10}\phi$ which, because the noise variance is set to 1, represents the SNR as well). Also, $\textbf{I}_{N_\text{r}}$ represents the ${N_\text{r}}\times {N_\text{r}}$ identity matrix.

Considering $T=1$ in (\ref{eq:outageprobinrfast}), the outage probability is rephrased as
\begin{align}\label{eq:outageprobinrslow}
\Pr(\text{Outage})^\text{Slow-fading}=\Pr\left(\frac{1}{M}\sum_{n=1}^{M}\log\left|\textbf{I}_{N_\text{r}}+\frac{\phi}{N_\text{t}}\textbf{H}(n)\textbf{H}(n)^h\right|\le \frac{R}{M}\right),
\end{align}
in a slow-fading channel. Also, setting $\textbf{H}(t)=\textbf{H},\forall t=1,\ldots,MT,$ the outage probability in a quasi-static fading channel is given by
\begin{align}\label{eq:outageprobinrslow}
\Pr(\text{Outage})^\text{Quasi-static}=\Pr\left(\log\left|\textbf{I}_{N_\text{r}}+\frac{\phi}{N_\text{t}}\textbf{H}\textbf{H}^h\right|\le \frac{R}{M}\right).
\end{align}
Using (\ref{eq:outageprobinrfast})-(\ref{eq:outageprobinrslow}) for given initial transmission rate and SNR, the problem formulation of the paper can be expressed as
\begin{align}\label{eq:problemformulation}
\{\hat N_\text{t},\hat N_\text{r}\}=\mathop {\arg\min }\limits_{N_\text{t},N_\text{r}}\{\Pr(\text{Outage})\le \theta\}.
\end{align}
Here, $\theta$ denotes an outage probability constraint and $\hat N_\text{t}, \hat N_\text{r}$ are the minimum numbers of transmit/receive antennas that are required to satisfy the outage probability constraint. In the following, we study (\ref{eq:problemformulation}) in four distinct cases:
\begin{itemize}
  \item Case 1: $N_\text{r}$ is large but $N_\text{t}$ is given.
  \item Case 2: $N_\text{r}$ is given but $N_\text{t}$ is large.
  \item Case 3: Both $N_\text{t}$ and $N_\text{r}$ are large and the transmission SNR is low.
  \item Case 4: Both $N_\text{t}$ and $N_\text{r}$ are large and the transmission SNR is high.
\end{itemize}
It is worth noting that the three first cases are commonly of interest in large MIMO systems. However, for the completeness of the discussions, we consider Case 4 as well. Moreover, in harmony with the literature \cite{6241389,6172680}\footnote{In \cite{6241389,6172680}, which study multi-user MIMO setups, $N_\text{t}$ and $N_\text{r}$ are supposed to follow (\ref{eq:ratioK}) while, as opposed to our work, they are considered to be asymptotically large.}, we analyze Cases 3-4 under the assumption
\begin{align}\label{eq:ratioK}
\frac{N_\text{t}}{N_\text{r}}=K,
\end{align}
with $K$ being a constant. However, as seen in the following, it is straightforward to extend the results of the paper to the cases with other relations between the numbers of antennas.
\section{Performance Analysis}
To solve (\ref{eq:problemformulation}), let us first introduce Lemma 1. The lemma is of interest because it represents the outage probability as a function of the number of antennas, and simplifies the performance analysis remarkably.

\textbf{\emph{Lemma 1:}} Considering Cases 1-4, the outage probability of the INR-based MIMO-HARQ system is given by
\begin{align}\label{eq:Qfuncoutageprobinrfast}
\left\{\begin{matrix}
\Pr(\text{Outage})^\text{Fast-fading}=Q\left(\frac{\sqrt{MT}(\mu-\frac{R}{M})}{\sigma}\right), & \text{(i)} \\
\Pr(\text{Outage})^\text{Slow-fading}=Q\left(\frac{\sqrt{M}(\mu-\frac{R}{M})}{\sigma}\right), & \text{(ii)} \\
\Pr(\text{Outage})^\text{Quasi-static}=Q\left(\frac{\mu-\frac{R}{M}}{\sigma}\right),\,\,\,\,\,\,\,\,\,\,\,\, &
\text{(iii)}\end{matrix}\right.
\end{align}
where $Q(x)=\frac{1}{\sqrt{2\pi}}\int_x^\infty{e^{-\frac{u^2}{2}}\text{d}u}$ is the Gaussian $Q$-function and for different cases $\mu$ and $\sigma$ are given in (\ref{eq:tarokhappxx}).
\begin{proof}
The proof is based on (\ref{eq:outageprobinrfast})-(\ref{eq:outageprobinrslow}) and \cite[Theorems 1-3]{1327795}, where considering Cases 1-4 the  random variable $Z(t)=\log|\textbf{I}_{N_\text{r}}+\frac{\phi}{N_\text{t}}\textbf{H}(t)\textbf{H}(t)^h|$ converges in distribution to a Gaussian random variable $Y\sim\mathcal{N}(\mu,\sigma^2)$ which, depending on the numbers of antennas, has the following characteristics
\vspace{-0mm}
\begin{align}\label{eq:tarokhappxx}
(\mu,\sigma^2)= \left\{ \begin{array}{l}
 \left(N_\text{t}\log\left(1+\frac{N_\text{r}\phi}{N_\text{t}}\right),\frac{N_\text{t}}{N_\text{r}}\right),\,\,\,\,\,\,\,\,\,\,\,\,\,\,\,\,\text{if Case 1} \\
 \left(N_\text{r}\log\left(1+\phi\right),\frac{N_\text{r}\phi^2}{N_\text{t}(1+\phi)^2}\right),\,\,\,\,\,\,\,\,\,\,\text{if Case 2} \\
\left(N_\text{r}\phi,\frac{N_\text{r}}{N_\text{t}}\phi^2\right),\,\,\,\,\,\,\,\,\,\,\,\,\,\,\,\,\,\,\,\,\,\,\,\,\,\,\,\,\,\,\,\,\,\,\,\,\,\,\,\,\,\,\text{if Case 3} \\
\left(\tilde\mu,\tilde\sigma^2\right),\,\,\,\,\,\,\,\,\,\,\,\,\,\,\,\,\,\,\,\,\,\,\,\,\,\,\,\,\,\,\,\,\,\,\,\,\,\,\,\,\,\,\,\,\,\,\,\,\,\,\,\,\,\,\,\,\text{if Case 4} \\
\tilde\mu=N_\text{min}\log\left(\frac{\phi}{N_\text{t}}\right)+N_\text{min}\left(\sum_{i=1}^{N_\text{max}-N_\text{min}}{\frac{1}{i}}-\gamma\right)+\sum_{i=1}^{N_\text{min}-1}{\frac{i}{N_\text{max}-i}},\,\gamma=0.5772\ldots\\
\tilde\sigma^2=\sum_{i=1}^{N_\text{min}-1}{\frac{i}{(N_\text{max}-N_\text{min}+i)^2}}+N_\text{min}\left(\frac{\pi^2}{6}-\sum_{i=1}^{N_\text{max}-1}{\frac{1}{i^2}}\right), \\N_\text{max}\doteq\max(N_\text{t},N_\text{r}),\,N_\text{min}\doteq\min(N_\text{t},N_\text{r}).\\
\end{array} \right. \\ \nonumber
\end{align}
In this way, from (\ref{eq:outageprobinrfast}) and for different cases, the outage probability in fast-fading condition is given by
\begin{align}
\Pr(\text{Outage})^\text{Fast-fading}=\Pr\left(Z\le \frac{R}{M}\right),Z\doteq\frac{1}{MT}\sum_{t=1}^{MT}{Z(t)},
\end{align}
where, because $Z$ is the average of $MT$ independent Gaussian random variables $Y\sim\mathcal{N}(\mu,\sigma^2)$, we have $Z\sim\mathcal{N}(\mu,\frac{1}{MT}\sigma^2)$. Consequently, using the cumulative distribution function (CDF) of the Gaussian random variables, the outage probability of the fast-fading condition is given by (\ref{eq:Qfuncoutageprobinrfast}.i). The same arguments can be applied to derive (\ref{eq:Qfuncoutageprobinrfast}.ii-iii) in the slow-fading and quasi-static conditions.
\end{proof}
Lemma 1 leads to the following corollaries:
\begin{itemize}
  \item[1)] For Cases 1-4, using INR MIMO-HARQ in the quasi-static, slow- and fast-fading channels leads to scaling the variance of the equivalent random variable by $1$, $M$ and $MT$, respectively. That is, using HARQ, there exists mappings between the quasi-static, the slow- and the fast-fading conditions in the sense that with proper scaling of $\sigma$ in (\ref{eq:Qfuncoutageprobinrfast}) they lead to the same outage probability.
  \item[2)] With asymptotically large numbers of transmit and/or receive antennas, the optimal data rate which leads to zero outage probability and maximum throughput is given by $R=\mu-\omega, \omega\to 0,$ with $\mu$ derived in (\ref{eq:tarokhappxx}); Interestingly, the result is independent of the fading condition. Also, with asymptotically high number of antennas and $R=\mu-\omega, \omega\to 0,$ no HARQ is needed because the message is decoded in the first round (with probability 1).
  \item[3)] Finally, using (\ref{eq:Qfuncoutageprobinrfast}), we can map the MIMO-HARQ system into an equivalent SISO-HARQ setup whose fading realizations follow $\mathcal{N}(\mu,\sigma^2)$ with $\mu$ and $\sigma$ given in (\ref{eq:tarokhappxx}) for different cases.
\end{itemize}
Using Lemma 1, the minimum numbers of antennas satisfying different outage probability constraints are determined as stated in Theorem 1.

\textbf{\emph{Theorem 1:}} The minimum numbers of the transmit and/or receive antennas in an INR-based MIMO-HARQ system that satisfy the outage probability constraint $\Pr(\text{Outage})\le \theta$ are given by
\begin{align}\label{eq:eqtheorem1}
\left\{\begin{matrix}
  \hat N_\text{r}=\frac{\left(Q^{-1}(\theta)\right)^2}{4MTN_\text{t}W^2\left(\frac{Q^{-1}(\theta)\sqrt{\phi}}{2\sqrt{MT}N_\text{t}}e^{-\frac{R}{2MN_\text{t}}}\right)}, \,\,\,\,\,\,\,\,\,\,\,\,\,\,\,\,\,\,\,\,\,\,\text{if Case 1}\\
  \hat N_\text{t}=\left(\frac{\phi \sqrt{N_\text{r}}Q^{-1}(\theta)}{\sqrt{MT}(1+\phi)\left(N_\text{r}\log(1+\phi)-\frac{R}{M}\right)}\right)^2,\,\,\,\,\,\,\,\,\,\,\,\,\,\,\,\,\,\,\,\text{if Case 2}\\
  \hat N_\text{r}=\hat N, \hat N_\text{t}=K\hat N, \hat N=\frac{R}{M\phi}+\frac{Q^{-1}(\theta)}{\sqrt{MTK}},\,\,\,\,\,\,\,\,\text{if Case 3}\\
  \hat N_\text{r}=\hat N, \hat N_\text{t}=K\hat N \,\,\,\,\,\,\,\,\,\,\,\,\,\,\,\,\,\,\,\,\,\,\,\,\,\,\,\,\,\,\,\,\,\,\,\,\,\,\,\,\,\,\,\,\,\,\,\,\,\,\,\,\,\,\,\,\,\,\text{if Case 4}\\
\left\{\begin{matrix}
\hat N\simeq\frac{\frac{R}{M}+\frac{Q^{-1}(\theta)}{\sqrt{MT}}\sqrt{\log\left(\frac{K}{K-1}\right)}}{\log(\phi)-\gamma-1+(K-1)\log\left(\frac{K}{K-1}\right)}, \,\,\,\,\,\,\,\,\,\,\,\,\,\,\,\,\,\,\,\,\,\,\,\,\,\,\,\,\, K>1\,\,\,\,\\
\hat N\simeq\frac{\frac{R}{M}+\frac{Q^{-1}(\theta)}{\sqrt{MT}}\sqrt{-\log({1-K})}}{K\left(\log(\phi)-\gamma-1-\log(K)+\left(\frac{K-1}{K}\right)\log(1-K)\right)}\,\,\,\,\, \,\,\, K< 1,
\end{matrix}\right.
\end{matrix}\right.
\end{align}
if the channel is fast-fading. Here, $Q^{-1}(x)$ and $W(x)$ denote the inverse $Q$-function and the Lambert W function, respectively. For the slow-fading and quasi-static conditions, the minimum numbers of the antennas are obtained by (\ref{eq:eqtheorem1}) where the term $\frac{Q^{-1}(\theta)}{\sqrt{MT}}$ is replaced by $\frac{Q^{-1}(\theta)}{\sqrt{M}}$ and ${Q^{-1}(\theta)}$, respectively.
\begin{proof}
Considering Lemma 1 and a fast-fading condition, (\ref{eq:problemformulation}) is rephrased as
\begin{align}\label{eq:fastproblemformulation2}
\{\hat N_\text{t},\hat N_\text{r}\}^\text{Fast-fading}=\mathop {\arg }\limits_{N_\text{t},N_\text{r}}\left\{\frac{\mu-\frac{R}{M}}{\sigma}=\frac{Q^{-1}(\theta)}{\sqrt{MT}}\right\},
\end{align}
which for different cases leads to

Case 1:
\begin{align}\label{eq:problemformulationcase1}
\hat N_\text{r}&=\mathop {\arg }\limits_{N_\text{r}}\left\{N_\text{t}\log\left(1+\frac{N_\text{r}\phi}{N_\text{t}}\right)-\frac{R}{M}=\frac{Q^{-1}(\theta)}{\sqrt{MT}}\sqrt{\frac{N_\text{t}}{N_\text{r}}}\right\}
\nonumber\\&
\mathop  \simeq \limits^{(a)}\frac{N_\text{t}}{\phi}\mathop {\arg }\limits_{u}\left\{\log(u)=\frac{R}{MN_\text{t}}+\frac{Q^{-1}(\theta)\sqrt{\phi}}{\sqrt{MT}N_\text{t}\sqrt{u}}\right\}\nonumber\\&\Rightarrow
\hat N_\text{r}=\frac{(Q^{-1}(\theta))^2}{4MTN_\text{t}W^2\left(\frac{Q^{-1}(\theta)\sqrt{\phi}}{2\sqrt{MT}N_\text{t}}e^{-\frac{R}{2MN_\text{t}}}\right)}.
\end{align}
Case 2:
\begin{align}\label{eq:problemformulationcase2}
\hat N_\text{t}&=\mathop {\arg }\limits_{N_\text{t}}\left\{N_\text{r}\log(1+\phi)-\frac{R}{M}=\frac{Q^{-1}(\theta)\phi \sqrt{N_\text{r}}}{\sqrt{MT}(1+\phi)\sqrt{N_\text{t}}}\right\}\nonumber\\&\Rightarrow \hat N_\text{t}=\left(\frac{\phi \sqrt{N_\text{r}}Q^{-1}(\theta)}{\sqrt{MT}(1+\phi)\left(N_\text{r}\log(1+\phi)-\frac{R}{M}\right)}\right)^2.
\end{align}
Case 3: $\hat N_\text{r}=\hat N, \hat N_\text{t}=K\hat N$ where
\begin{align}\label{eq:problemformulationcase3}
\hat N&=\mathop {\arg }\limits_{N}\left\{N\phi-\frac{R}{M}=\frac{Q^{-1}(\theta)\phi}{\sqrt{MTK}}\right\}\Rightarrow \hat N=\frac{R}{M\phi}+\frac{Q^{-1}(\theta)}{\sqrt{MTK}}.
\end{align}
In (\ref{eq:problemformulationcase1}), $(a)$ is obtained by using the approximation $\log(1+u)\simeq\log(u)$ for large $u$'s and variable transform $u=\frac{N_\text{r}\phi}{N_\text{t}}.$ Also, $W(x)$ denotes the Lambert W function defined as $ye^y=x\Rightarrow y=W(x)$ \cite{Lambertw}. Note that the Lambert W function has an efficient implementation in MATLAB and MATHEMATICA.

For Case 4, we consider two scenarios and use the following approximations.

Case 4 with $K>1$: Then, $\hat N_\text{r}=\hat N, \hat N_\text{t}=K\hat N$ and
\begin{align}\label{eq:problemformulationcase41}
\hat N&=\mathop {\arg }\limits_{N}\Bigg\{N\log\left(\frac{\phi}{KN}\right)+N\left(\sum_{i=1}^{(K-1)N}{\frac{1}{i}}-\gamma\right)+\sum_{i=1}^{N-1}{\frac{i}{KN-i}}-\frac{R}{M}\nonumber\\&\,\,\,\,\,\,\,\,\,\,\,\,\,\,\,\,\,\,\,\,\,\,\,\,\,\,\,\,\,\,\,\,\,\,\,\,\,\,\,\,\,\,\,\,\,\,\,\,\,\,=\frac{Q^{-1}(\theta)}{\sqrt{MT}}\sqrt{\sum_{i=1}^{N-1}{\frac{i}{\left((K-1)N+i\right)^2}}+N\left(\frac{\pi^2}{6}-\sum_{i=1}^{KN-1}{\frac{1}{i^2}}\right)}\Bigg\}
\nonumber\\&\mathop  \simeq \limits^{(b)} \mathop {\arg }\limits_{N}\Bigg\{N\log\left(\frac{\phi}{KN}\right)+N\left(\log((K-1)N)-\gamma\right)+2-N+KN\log\left(\frac{KN-1}{(K-1)N+1}\right)-\frac{R}{M}\nonumber\\&\,\,\,\,\,\,\,\,\,\,\,\,\,\,\,\,\,\,=\frac{Q^{-1}(\theta)}{\sqrt{MT}}\sqrt{\frac{(K-1)N(2-N)}{(KN-N+1)(KN-1)}+\log\left(\frac{KN-1}{(K-1)N+1}\right)+N\left(\frac{\pi^2}{6}-\sum_{i=1}^{KN-1}{\frac{1}{i^2}}\right)}\Bigg\}
\nonumber\\&\mathop  \simeq \limits^{(c)}\mathop {\arg }\limits_{N}\Bigg\{N\left(\log(\phi)-\gamma-1+(K-1)\log\left(\frac{K}{K-1}\right)\right)-\frac{R}{M}=\frac{Q^{-1}(\theta)}{\sqrt{MT}}\sqrt{\log\left(\frac{K}{K-1}\right)}\Bigg\}\nonumber\\&\Rightarrow \hat N=\frac{\frac{R}{M}+\frac{Q^{-1}(\theta)}{\sqrt{MT}}\sqrt{\log\left(\frac{K}{K-1}\right)}}{\log(\phi)-\gamma-1+(K-1)\log\left(\frac{K}{K-1}\right)}.
\end{align}
Here, $(b)$ is obtained by implementing the Riemann integral approximation  $\sum_{i=1}^{n}{f(i)}\simeq\int_1^n{f(x)\text{d}x}$ in the first three summation terms. Then, $(c)$ follows from some manipulations, the fact that $N$ is assumed large, and $N(\frac{\pi^2}{6}-\sum_{i=1}^{KN}{\frac{1}{i^2}})\to\frac{1}{K}$ for large $N$'s.

Case 4 with $K<1$: Then, $\hat N_\text{r}=\hat N, \hat N_\text{t}=K\hat N$ and
\begin{align}\label{eq:problemformulationcase42}
&\hat N=\mathop {\arg }\limits_{N}\Bigg\{KN\log\left(\frac{\phi}{KN}\right)+KN\left(\sum_{i=1}^{(1-K)N}{\frac{1}{i}}-\gamma\right)+\sum_{i=1}^{KN-1}{\frac{i}{N-i}}-\frac{R}{M}\nonumber\\&\,\,\,\,\,\,\,\,\,\,\,\,\,\,\,\,\,\,\,\,\,\,\,\,\,\,\,\,\,\,\,\,\,\,\,\,\,\,\,\,\,\,\,\,\,\,\,\,\,\,=\frac{Q^{-1}(\theta)}{\sqrt{MT}}\sqrt{\sum_{i=1}^{KN-1}{\frac{i}{((1-K)N+i)^2}}+KN\left(\frac{\pi^2}{6}-\sum_{i=1}^{N-1}{\frac{1}{i^2}}\right)}\Bigg\}
\nonumber\\&\mathop  \simeq \limits^{(b)} \mathop {\arg }\limits_{N}\Bigg\{KN\log\left(\frac{\phi}{KN}\right)+KN\left(\log((1-K)N)-\gamma\right)+2-KN+N\log\left(\frac{N-1}{(1-K)N+1}\right)-\frac{R}{M}\nonumber\\&\,\,\,\,\,\,\,\,\,\,\,\,\,\,\,\,\,\,\,\,\,\,=\frac{Q^{-1}(\theta)}{\sqrt{MT}}\sqrt{\frac{(1-K)N(2-NK)}{(N-NK+1)(N-1)}+\log\left(\frac{N-1}{(1-K)N+1}\right)+NK\left(\frac{\pi^2}{6}-\sum_{i=1}^{N-1}{\frac{1}{i^2}}\right)}\Bigg\}
\nonumber\\&\mathop  \simeq \limits^{(c)}\mathop {\arg }\limits_{N}\Bigg\{NK\bigg(\log(\phi)-\gamma-1-\log(K)+\left(\frac{K-1}{K}\right)\log(1-K)\bigg)-\frac{R}{M}=\frac{Q^{-1}(\theta)}{\sqrt{MT}}\sqrt{-\log({1-K})}\Bigg\}\nonumber\\&\Rightarrow \hat N=\frac{\frac{R}{M}+\frac{Q^{-1}(\theta)}{\sqrt{MT}}\sqrt{-\log({1-K})}}{K\left(\log(\phi)-\gamma-1-\log(K)+(\frac{K-1}{K})\log(1-K)\right)},
\end{align}
where $(b)$ and $(c)$ are obtained with the same procedure as in (\ref{eq:problemformulationcase41})\footnote{We can follow the same procedure as in (\ref{eq:problemformulationcase41})-(\ref{eq:problemformulationcase42}) to write
\begin{align}\label{eq:sagK1}
\hat N&=\mathop {\arg }\limits_{N}\Bigg\{N(\log(\phi)-\gamma-1)-\frac{R}{M}=\frac{Q^{-1}(\theta)}{\sqrt{MT}}\sqrt{\log(N-1)+1}\Bigg\},
\end{align}
in the cases with $K=1$ which can be solved numerically via, e.g., ``'fsolve'' function of MATLAB or by different approximation schemes. However, for simplicity and because it is a special condition, we do not consider $K=1$ as a separate case.}. Finally, note that with slow-fading and quasi-static fading channel (\ref{eq:fastproblemformulation2}) is rephrased as
\begin{align}\label{eq:slowproblemformulation2}
\{\hat N_\text{t},\hat N_\text{r}\}^\text{Slow-fading}=\mathop {\arg }\limits_{N_\text{t},N_\text{r}}\left\{\frac{\mu-\frac{R}{M}}{\sigma}=\frac{Q^{-1}(\theta)}{\sqrt{M}}\right\}
\end{align}
and
\begin{align}\label{eq:quasiproblemformulation2}
\{\hat N_\text{t},\hat N_\text{r}\}^\text{Quasi-static}=\mathop {\arg }\limits_{N_\text{t},N_\text{r}}\left\{\frac{\mu-\frac{R}{M}}{\sigma}={Q^{-1}(\theta)}\right\},
\end{align}
respectively. Therefore, as stated in the theorem, with slow-fading and quasi-static channel the required numbers of the transmit and/or receive antennas are determined by (\ref{eq:eqtheorem1}) while the term $\frac{Q^{-1}(\theta)}{\sqrt{MT}}$ is replaced by $\frac{Q^{-1}(\theta)}{\sqrt{M}}$ and ${Q^{-1}(\theta)}$, respectively.
\end{proof}

According to Theorem 1, the following conclusions can be drawn:
\begin{itemize}
  \item[1)] Using the tight approximation $W(e^{a+x})\simeq x+a-\log(a+x)$ in (\ref{eq:eqtheorem1}), the required number of receive antennas in Case 1 is rephrased as
      \begin{align}\label{eq:approxcase1}
      \hat N_\text{r}&\simeq\frac{\left(Q^{-1}(\theta)\right)^2}{4MTN_\text{t}\left(\log\left(\frac{Q^{-1}(\theta)\sqrt{\phi}}{2\sqrt{MT}N_\text{t}}\right)-\frac{R}{2MN_\text{t}}-\log\left(\log\left(\frac{Q^{-1}(\theta)\sqrt{\phi}}{2\sqrt{MT}N_\text{t}}\right)-\frac{R}{2MN_\text{t}}\right)\right)^2}\nonumber\\&\simeq \frac{\left(Q^{-1}(\theta)\right)^2}{MTN_\text{t}\left(\log({\phi})\right)^2},
      \end{align}
      where the last approximation holds for moderate/high SNRs. Thus, at moderate/high SNR regimes, the required number of receive antennas increases with $(Q^{-1}(\theta))^2$ linearly. On the other hand, the required number of receive antennas is inversely proportional to the number of experienced fading realizations $MT,$ the number of transmit antennas $N_\text{t}$, and $(\log(\phi))^2$. Interestingly, we can use (\ref{eq:eqtheorem1}.Case 2) to show that at high SNRs the same scaling laws hold for Cases 1 and 2. That is, in Case 2, the required number of transmit antennas decreases (resp. increases) with $MT,$ $N_\text{t}$, and $(\log(\phi))^2$ (resp. $(Q^{-1}(\theta))^2$) linearly.
      %
  \item[2)] The same scaling laws are valid in Cases 3 and 4, i.e., when the numbers of transmit and receive antennas increase simultaneously. For instance, the required number of antennas increases with $Q^{-1}(\theta)$ and the code rate $R$ semi-linearly\footnote{The variable $y$ is semilinear with $x$ if $y=a+bx$ for given constants $a$ and $b.$} (see (\ref{eq:eqtheorem1}.Cases 3-4)). At hard outage probability constraints, i.e., small values of $\theta$, the required number of antennas decreases with the number of retransmissions according to $\frac{1}{\sqrt{M}}.$ On the other hand, the number of antennas decreases with $M$ linearly when the outage constraint is relaxed, i.e., $\theta$ increases. The only difference between Cases 3 and 4 is that in Case 3 (resp. Case 4) the number of antennas decreases with $\phi$ (resp. $\log(\phi)$) linearly.
  \item[3)] It has been previously proved that at low SNRs the same performance is achieved by the MIMO systems using INR and repetition time diversity (RTD) HARQ \cite[Section V.B]{MIMOARQkhodemun}. Thus, although the paper concentrates on the INR HARQ, the same number of antennas are required in the MIMO-RTD setups, as long as the SNR is low.
\end{itemize}

As the number of antennas increases, the CDF of the accumulated mutual information, e.g., $\frac{1}{MT}\sum_{n=1}^{M}\sum_{t=(n-1)T+1}^{nT}\log|\textbf{I}_{N_\text{r}}+\frac{\phi}{N_\text{t}}\textbf{H}(t)\textbf{H}(t)^h|$ in fast-fading conditions, tends towards the step function.
Therefore, depending on the SNR and the initial rate, the outage probability rapidly converges to either zero or one as the number of transmit and/or receiver antennas increases.  To further elaborate on this point and investigate the effect of the number of antennas, we define the normalized outage factor as
\begin{align}\label{eq:eqdefnormalizedoutage1}
\Gamma=-\frac{\log(\Pr(\text{Outage}))}{N_\text{t}N_\text{r}}.
\end{align}
Intuitively, (\ref{eq:eqdefnormalizedoutage1}) gives the negative of the slope of the outage probability curve plotted versus the product of the numbers of transmit/receive antennas. Also, (\ref{eq:eqdefnormalizedoutage1}) follows the same concept as in the diversity gain $D=-\lim_{\phi\to\infty}\frac{\log(\Pr(\text{Outage}))}{\phi}$ \cite[Eq. 14]{a01661837} which is an efficient metric for the asymptotic analysis of the MIMO setups. Theorem 2 studies the normalized outage factor in more details.

\textbf{\emph{Theorem 2:}} For Cases 1-4, different fading conditions and appropriate initial rates/SNR, the normalized outage factor is approximated by (\ref{eq:eqdefnormalizedoutagefast1}).

\begin{proof}
With given initial transmission rate $R$ and SNR $\phi$, we use the approximation
\begin{align}\label{eq:Qfuncappr}
Q(x)\simeq\frac{e^{-\frac{x^2}{2}}}{2}, x\ge 0,
\end{align}
for large $x$'s and (\ref{eq:Qfuncoutageprobinrfast}) to rewrite the normalized outage factor (\ref{eq:eqdefnormalizedoutage1}) as
\begin{align}\label{eq:eqdefnormalizedoutagefast1}
&\Gamma=\frac{c\left(\mu-\frac{R}{M}\right)^2}{2N_\text{t}N_\text{r}\sigma^2},\nonumber\\&
c=\left\{\begin{matrix}
MT,\,\, \text{ Fast-fading}\,\,\,\\
M, \,\,\,\,\,\text{ Slow-fading}\\
1, \,\,\,\,\,\,\,\,\,\text{   Quasi-static.}
\end{matrix}\right.
\end{align}
Then, from (\ref{eq:tarokhappxx}), the normalized outage factor in different cases is found as
\begin{align}\label{eq:eqdefnormalizedoutagefast1}
\Gamma=
\left\{\begin{matrix}
 \frac{c}{2N_\text{t}N_\text{r}}\frac{\left(N_\text{t}\log\left(1+\frac{N_\text{r}\phi}{N_\text{t}}\right)-\frac{R}{M}\right)^2}{\frac{N_\text{t}}{N_\text{r}}}=\frac{c}{2}\left(\log\left(1+\frac{N_\text{r}\phi}{N_\text{t}}\right)-\frac{R}{MN_\text{t}}\right)^2\,\,\,\,\,\,\,\,\,\,\,\,\,\,\,\,\,\,\,\,\,\,\,\,\,\text{ if Case 1}\\
\frac{c}{2N_\text{t}N_\text{r}}\frac{\left(N_\text{r}\log(1+\phi)-\frac{R}{M}\right)^2N_\text{t}(1+\phi)^2}{N_\text{r}\phi^2}=\frac{c}{2}\frac{(1+\phi)^2}{\phi^2}\left(\log(1+\phi)-\frac{R}{MN_\text{r}}\right)^2 \,\,\,\,\,\,\,\,\, \text{if Case 2}\\
\frac{c}{2KN^2}\frac{\left(N\phi-\frac{R}{M}\right)^2}{\frac{\phi^2}{K}}=\frac{c}{2}\,\,\,\,\,\,\,\,\,\,\,\,\,\,\,\,\,\,\,\,\,\,\,\,\,\,\,\,\,\,\,\,\,\,\,\,\,\,\,\,\,\,\,\,\,\,\,\,\,\,\,\,\,\,\,\,\,\,\,\,\,\,\,\,\,\,\,\,\,\,\,\,\,\,\,\,\,\,\,\,\,\,\,\,\,\,\,\,\,\,\,\,\,\,\,\,\,\,\,\,\,\,\,\,\,\,\,\,\,\, \text{if Case 3}\\
\frac{c}{2K}\alpha,\,\,\,\,\,\,\,\,\,\,\,\,\,\,\,\,\,\,\,\,\,\,\,\,\,\,\,\,\,\,\,\,\,\,\,\,\,\,\,\,\,\,\,\,\,\,\,\,\,\,\,\,\,\,\,\,\,\,\,\,\,\,\,\,\,\,\,\,\,\,\,\,\,\,\,\,\,\,\,\,\,\,\,\,\,\,\,\,\,\,\,\,\,\,\,\,\,\,\,\,\,\,\,\,\,\,\,\,\,\,\,\,\,\,\,\,\,\,\,\,\,\,\,\,\,\,\,\,\,\,\,\,\,\,\,\,\,\,\,\,\,\,\,\,\,\text{if Case 4} \\
\alpha=\left\{\begin{matrix}
\frac{\left(\log(\phi)+(K-1)\log\left(\frac{K}{K-1}\right)-\gamma-1\right)^2}{\log\left(\frac{K}{K-1}\right)}\text{ if } K>1\\
\frac{K^2\left(\log(\phi)+\frac{K-1}{K}\log(1-K)-\log(K)-\gamma -1\right)^2}{-\log(1-K)}\text{  if } K< 1,
\end{matrix}\right.\,\,\,\,\,\,\,\,\,\,\,\,\,\,\,\,\,\,\,\,\,\,\,\,\,\,\,\,\,\,\,\,\,\,\,\,\,\,\,\,\,\,
\\
c=\left\{\begin{matrix}
MT,\,\, \text{ Fast-fading}\,\,\,\\
M, \,\,\,\,\,\text{ Slow-fading}\\
1, \,\,\,\,\,\,\,\,\,\text{   Quasi-static,}
\end{matrix}\right.\,\,\,\,\,\,\,\,\,\,\,\,\,\,\,\,\,\,\,\,\,\,\,\,\,\,\,\,\,\,\,\,\,\,\,\,\,\,\,\,\,\,\,\,\,\,\,\,\,\,\,\,\,\,\,\,\,\,\,\,\,\,\,\,\,\,\,\,\,\,\,\,\,\,\,\,\,\,\,\,\,\,\,\,\,\,\,\,\,\,\,\,\,\,\,\,\,\,\,\,\,\,\,
\end{matrix}\right.
\end{align}
where $\alpha$ is found by following the same approach as in (\ref{eq:problemformulationcase41})-(\ref{eq:problemformulationcase42}). Finally, note that to use (\ref{eq:Qfuncappr}) the initial rate and the SNR should be such that $\mu\ge \frac{R}{M}$ for the considered number of antennas. Otherwise, the outage probability converges to 1 and $\Gamma\to 0.$
\end{proof}
Interestingly, the theorem indicates that:
\begin{itemize}
  \item[1)] The normalized outage factor becomes constant in all cases, except Case 1 with a given (resp. large) number of transmit (resp. receive) antennas. Intuitively, this is because in all cases (except Case 1) the power per transmit antenna decreases by increasing the number of transmit antennas. Therefore, there is a tradeoff between increasing the diversity and reducing the power per antenna and, as a result, the normalized outage factor converges to the values given in (\ref{eq:eqdefnormalizedoutagefast1}). In Case 1, however, the message decoding probability is always increased by increasing the number of receive antennas and, as seen in Theorem 2, the normalized outage factor increases with $N_\text{r}$ monotonically, as long as $\mu\ge \frac{R}{M}$.
  \item[2)] In Case 3, the normalized outage factor becomes independent of the transmission SNR as long as $\mu\ge \frac{R}{M}$. In Cases 1, 2 and 4, on the other hand, the normalized outage factor scales with the SNR according to $(\log(\phi))^2,$ if the SNR is high.
  \item[3)] In all cases, the normalized outage factor scales with the number of experienced fading realizations during the packet transmission, i.e., $MT,$ linearly. Note that the same conclusion has been previously derived for the diversity gain $D=-\lim_{\phi\to\infty}\frac{\log(\Pr(\text{Outage}))}{\phi}$ \cite{a04608960,a06035974}.
  \item[4)] In cases 3-4, the normalized outage factor does not depend on the initial transmission rate. Moreover, in Case 3 the normalized outage factor is independent of the ratio between the number of transmit and receive antennas.
\end{itemize}

\subsection{On the Effect of Power Amplifiers}
As the number of the transmit antennas increases, it is important to take the efficiency of radio-frequency PAs into account \cite{6736761,6798744,6375940}. For this reason, we use Lemma 1 to investigate the system performance in the cases with non-ideal PAs as follows.

It has been previously shown that the PA efficiency can be written as \cite{phdthesisBjornemo,4160747}, \cite[Eq. (3)]{6515206} and \cite[Eq. (3)]{6725577}
\begin{align}\label{eq:ampmodeldaniel}
&\frac{\phi}{\phi^\text{cons}}=\epsilon\left(\frac{\phi}{\phi^\text{max}}\right)^\vartheta\,\Rightarrow  \phi=\sqrt[1-\vartheta]{\frac{\epsilon \phi^\text{cons}}{(\phi^\text{max})^\vartheta}}.
\end{align}
Here, $\phi, \phi^\text{max}$ and $\phi^\text{cons}$ are the output, the maximum output, and the consumed power of the PA, respectively, $\epsilon\in [0,1]$ denotes the maximum power efficiency achieved at $\phi=\phi^\text{max}$, and $\vartheta$ is a parameter that, depending on the PA classes, varies between $[0,1]$.
In this way, and because the INR-based MIMO-HARQ setup can be mapped into an equivalent SISO-HARQ system (see Lemma 1 and its following discussions), the equivalent mean and variances (\ref{eq:tarokhappxx}) are rephrased as
\vspace{-0mm}
\begin{align}\label{eq:PAtarokhappxx}
(\mu,\sigma^2)= \left\{ \begin{array}{l}
 \left(N_\text{t}\log\left(1+\frac{N_\text{r}}{N_\text{t}}\sqrt[1-\vartheta]{\frac{\epsilon \phi^\text{cons}}{(\phi^\text{max})^\vartheta}}\right),\frac{N_\text{t}}{N_\text{r}}\right),\,\,\,\,\,\,\,\,\,\,\,\,\,\,\,\,\,\,\,\,\,\,\,\,\,\,\,\,\,\,\,\,\,\,\,\,\,\text{Case 1} \\
 \left(N_\text{r}\log\left(1+\sqrt[1-\vartheta]{\frac{\epsilon \phi^\text{cons}}{(\phi^\text{max})^\vartheta}}\right),\frac{N_\text{r}}{N_\text{t}\left(1+\sqrt[1-\vartheta]{\frac{(\phi^\text{max})^\vartheta}{\epsilon \phi^\text{cons}}}\right)^2}\right),\,\,\text{Case 2} \\
\left(N_\text{r}\sqrt[1-\vartheta]{\frac{\epsilon \phi^\text{cons}}{(\phi^\text{max})^\vartheta}},\frac{N_\text{r}}{N_\text{t}}\sqrt[1-\vartheta]{\left(\frac{\epsilon \phi^\text{cons}}{(\phi^\text{max})^\vartheta}\right)^2}\right),\,\,\,\,\,\,\,\,\,\,\,\,\,\,\,\,\,\,\,\,\,\,\,\,\,\,\,\,\,\,\,\,\,\,\,\,\,\,\,\text{Case 3} \\
(\tilde\mu,\tilde\sigma^2),\,\,\,\,\,\,\,\,\,\,\,\,\,\,\,\,\,\,\,\,\,\,\,\,\,\,\,\,\,\,\,\,\,\,\,\,\,\,\,\,\,\,\,\,\,\,\,\,\,\,\,\,\,\,\,\,\,\,\,\,\,\,\,\,\,\,\,\,\,\,\,\,\,\,\,\,\,\,\,\,\,\,\,\,\,\,\,\,\,\,\,\,\,\,\,\,\,\,\,\,\,\,\,\,\,\text{Case 4} \\
\tilde\mu=N_\text{min}\log\left(\frac{1}{N_\text{t}}\sqrt[1-\vartheta]{\frac{\epsilon \phi^\text{cons}}{(\phi^\text{max})^\vartheta}}\right)+N_\text{min}\left(\sum_{i=1}^{N_\text{max}-N_\text{min}}{\frac{1}{i}}-\gamma\right)+\sum_{i=1}^{N_\text{min}-1}{\frac{i}{N_\text{max}-i}},\\
\tilde\sigma^2=\sum_{i=1}^{N_\text{min}-1}{\frac{i}{(N_\text{max}-N_\text{min}+i)^2}}+N_\text{min}\left(\frac{\pi^2}{6}-\sum_{i=1}^{N_\text{max}-1}{\frac{1}{i^2}}\right), \\N_\text{max}=\max(N_\text{t},N_\text{r}),\,N_\text{min}=\min(N_\text{t},N_\text{r}),\, \gamma=0.5772\ldots.\\
\end{array} \right. \\ \nonumber
\end{align}
in the cases with non-ideal PAs. This is the only modification required for the non-ideal PA scenario and the rest of the analysis remains the same as before.
\vspace{-0mm}
\subsection{On the Effect of Power Allocation}
Throughout the paper, we studied the system performance assuming a peak power constraint at the transmitter. However, the system performance is improved if the transmission powers are updated in the HARQ retransmission rounds.

Let the transmission power in the $m$th round be $\phi_m$. Then, the outage probability  in the fast-fading condition\footnote{For simplicity, the results of this part are given mainly for the fast-fading condition. It is straightforward to extend the results to the cases with other fading models.}, i.e., (\ref{eq:outageprobinrfast}), is rephrased as
\begin{align}\label{eq:outageprobinrfastopt}
\Pr(\text{Outage})^\text{Fast-fading}&=\Pr\left(\frac{1}{MT}\sum_{m=1}^{M}\sum_{t=(m-1)T+1}^{mT}\log\left|\textbf{I}_{N_\text{r}}+\frac{\phi_m}{N_\text{t}}\textbf{H}(t)\textbf{H}(t)^h\right|\le \frac{R}{M}\right)\nonumber\\&\mathop  = \limits^{(d)} \Pr\left(\frac{1}{MT}\sum_{m=1}^M{ Z_m}\le \frac{R}{M}\right)\mathop  = \limits^{(e)}Q\left(\frac{\bar \mu_{(M)}-\frac{R}{M}}{\bar\sigma_{(M)}}\right),\nonumber\\&
\bar \mu_{(m)}=\frac{1}{m}\sum_{n=1}^m{\mu_n},
\bar\sigma_{(m)}^2=\frac{1}{Tm^2}\sum_{n=1}^m{\sigma_n^2},
\end{align}
where $\mu_n$ and $\sigma_n$ are obtained by replacing $\phi_n$ into (\ref{eq:tarokhappxx}). Here, $(d)$ is obtained by $ Z_m\doteq \sum_{t=(m-1)T+1}^{mT}\log|\textbf{I}_{N_\text{r}}+\frac{\phi_m}{N_\text{t}}\textbf{H}(t)\textbf{H}(t)^h|\sim\mathcal{N}(T\mu_m,T\sigma_m^2)$. Also, $(e)$ is based on the fact that the sum of independent Gaussian random variables is a Gaussian random variable with the mean and variance equal to the sum of the variables means and variances, respectively.

If the message is correctly decoded in the $m$th round, the total transmission energy and the total number of channel uses are $\xi_{(m)}=L\sum_{n=1}^m{\phi_n}$ and $l_{(m)}=mL$, respectively. Also, the total transmission energy and the number of channel uses are $\xi_M=L\sum_{n=1}^M{\phi_M}$ and $l_{(M)}=ML$ if an outage occurs, where all possible retransmission rounds are used. Thus, we can follow the same procedure as in \cite{Tcomkhodemun,MIMOARQkhodemun,throughputdef} to find the average power, defined as the expected transmission energy over the expected number of channel uses, as
\begin{align}\label{eq:eqphibar}
\bar \Phi&= \frac{\phi_1+\sum_{m=1}^{M-1}{\phi_{m+1}\Pr\left(\frac{1}{Tm}\sum_{n=1}^{m}\sum_{t=(n-1)T+1}^{nT}\log\left|\textbf{I}_{N_\text{r}}+\frac{\phi_n}{N_\text{t}}\textbf{H}(t)\textbf{H}(t)^h\right|\le \frac{R}{m}\right)}}{1+\sum_{m=1}^{M-1}{\Pr\left(\frac{1}{Tm}\sum_{n=1}^{m}\sum_{t=(n-1)T+1}^{nT}\log\left|\textbf{I}_{N_\text{r}}+\frac{\phi_n}{N_\text{t}}\textbf{H}(t)\textbf{H}(t)^h\right|\le \frac{R}{m}\right)}}\nonumber\\&=
\frac{\phi_1+\sum_{m=1}^{M-1}{\phi_{m+1}Q\left(\frac{\bar \mu_{(m)}-\frac{R}{m}}{\bar\sigma_{(m)}}\right)}}{1+\sum_{m=1}^{M-1}{Q\left(\frac{\bar \mu_{(m)}-\frac{R}{m}}{\bar\sigma_{(m)}}\right)}}.
\end{align}
In this way, with a power constraint $\bar \Phi\le \phi,$ the problem formulation (\ref{eq:problemformulation}) is rephrased as
\begin{align}\label{eq:problemformulationopt}
\{\hat N_\text{t},\hat N_\text{r}\}=&\mathop {\arg\min }\limits_{N_\text{t},N_\text{r}}\{\Pr(\text{Outage})\le \theta\}
\nonumber\\&
\text{s.t.} \bar \Phi\le\phi,
\end{align}
which, using (\ref{eq:tarokhappxx}) and (\ref{eq:outageprobinrfastopt}), can be solved numerically or analytically.

\section{Simulation Results and Discussions}

In this section, we verify the accuracy of the derived results, and present the simulation results in spatially independent and correlated fading conditions as follows.

\subsection{Performance Analysis in Spatially-independent Fading Conditions}

In Figs. 1-4, we verify the accuracy of the results in Theorem 1 and derive the required number of transmit/receive antennas in the outage-limited conditions.  Setting $M=2, N_\text{t}=1$ (Case 1), and the outage probability constraints $\Pr(\text{Outage})\le\theta$ (with $\theta= 10^{-4}, 10^{-2}),$ Fig. 1 shows the required number of receive antennas versus the initial transmission rate $R$. The results of the figure are obtained for the slow-fading conditions and different transmission SNRs. Then, considering $N_\text{r}=1$ or $2,$ Fig. 2 demonstrates the required number of transmit antennas in Case 2 with large $N_\text{t}$ and given $N_\text{r}.$ Here, we consider quasi-static, slow- and fast-fading conditions with $\theta=10^{-4}, T=2, M=2, \phi=15 \text{ dB.}$ In Fig. 3, we verify the effect of HARQ on the system performance. Here, assuming Case 1 (large $N_\text{r}$ and $N_\text{t}=1,5$), the required number of antennas is derived in the scenarios with ($M=2$) and without ($M=1$) HARQ. The results of the figure are given for quasi-static channels, $\phi=5 \text{ dB}$ and  $\theta=10^{-4}.$

Figure 4 studies the required number of antennas in Cases 3 and 4 with low and high SNRs, respectively, large number of transmit and receive antennas, and $\frac{N_\text{t}}{N_\text{r}}=K$. Also, the figure demonstrates the analytical results of Theorem 1 when the approximation steps $(b)-(c)$ of (\ref{eq:problemformulationcase41})-(\ref{eq:problemformulationcase42}) are not implemented, i.e., (\ref{eq:fastproblemformulation2}) is solved numerically via (\ref{eq:tarokhappxx}). Here, we consider the quasi-static conditions, $M=1$, and $\theta=10^{-3}.$ Note that, to have the simulation results of Case 4 in reasonable running time, we have stopped the simulations at moderate initial transmission rates. For this reason, the simulation results of Case 4, i.e., the red solid-line curves of Case 4 in Fig. 4, are plotted for the moderate initial rates.

In Fig. 5, we analyze the normalized outage factor and evaluate the theoretical results of Theorem 2. Considering quasi-static conditions, $M=1, N_\text{t}=1$ and $\phi=5\text{dB,}$ Fig. 5a shows the outage probability versus the product of the number of transmit and receive antennas. Also, Fig. 5b demonstrates the normalized outage factor in Case 1 and compares the results with the theoretical derivations of Theorem 2. Finally, Fig. 5c studies the outage probability in Case 2 and compares the slope of the curves with the normalized outage factor derived in Theorem 2. Here, the results are obtained for the slow-  and fast-fading conditions ($T=2$) with $R=1, M=1, N_\text{r}=1$ and $\phi=5\text{dB}.$

Figure 6 evaluates the effect of non-ideal PAs and adaptive power allocation on the performance of large MIMO setups. Considering fast-fading conditions with $T=2$, Case 2 with large (resp. given) number of transmit (resp. receive) antennas and the outage probability constraint $\Pr(\text{Outage})\le \theta,\theta=10^{-4},$ Fig. 6a demonstrates the supported initial transmission rates, i.e., the maximum rates for which the outage probability is guaranteed, versus the total consumed power. For the non-ideal PA, we set $\phi^{\text{max}}=30\text{ dB}, \vartheta=0.5, \epsilon=0.65$, while the ideal PA corresponds to $\phi^{\text{max}}\to\infty, \vartheta=0, \epsilon=1$ in (\ref{eq:ampmodeldaniel}). The figure demonstrates the simulation results while, with the parameter settings of the figure, the same (with high accuracy) results are obtained if the supported initial rates are derived analytically according to (\ref{eq:PAtarokhappxx}) (Also, see Fig. 2 for the tightness of approximations in Case 2). Finally, assuming slow-fading conditions and Case 2 with $N_\text{r}=1,$ $\theta=10^{-3}, M = 2$,  Fig. 6b compares the  required number of transmit antennas in the scenarios with and without adaptive power allocation between the HARQ retransmissions. Note that, in all cases, we have also investigated a wider range of parameters and fading conditions, but because the performances of those cases have followed the same trends as the ones
shown, we have not included those results to avoid unnecessary complexity. According to the results, the following conclusions can be drawn:
\begin{itemize}
  \item For Cases 1-3 and different fading conditions, the analytical results of Theorem 1 are very tight for a broad range of initial transmission rates, outage probability constraints and SNRs (Figs. 1-3). Also, in Case 1 (resp. Case 2) the tightness of the approximations increases with the number of receive (resp. transmit) antennas (Figs. 1-2). Moreover, the approximation scheme of Theorem 1 can accurately determine the required number of antennas in Case 3 with different values of $K.$ For Case 4 (which is not of practical interest in large MIMO setups), we can find the required number of antennas accurately through Theorem 1 when (\ref{eq:fastproblemformulation2}) is solved numerically via (\ref{eq:tarokhappxx}). As such, the approximations $(b)-(c)$ of (\ref{eq:problemformulationcase41})-(\ref{eq:problemformulationcase42})  decrease the accuracy, although the curves still follow the same trend as in the simulation results. For instance, with different approximation approaches of Case 4, the required number of antennas increases with the initial rate linearly, in harmony with the simulation results (Fig. 4). The tightness of the approximations in Cases 3 (resp. Case 4) increases when the SNR decreases (resp. increases). Finally, the scaling laws of Theorem 1 are valid because, as demonstrated in Figs. 1-4, in all cases the analytical and the simulation results follow the same trends (see Theorem 1 and its following discussions).
  \item In all Cases, better approximation is achieved via Theorem 1 in fast-fading (resp. slow-fading) conditions compared to slow-fading (resp. quasi-static) conditions. This is intuitively because the central limit Theorem provides better approximation in Lemma 1 when the number of experienced fading realizations increases.
  \item The required number of antennas decreases as the outage probability constraint is relaxed, i.e., $\theta$ increases, while for different transmission SNRs, there is (almost) a fixed gap between the curves of different outage probability constraints (Fig. 1). Also, fewer antennas are required when the number of fading realizations experienced during the HARQ packet transmission increases. Intuitively, this is because more diversity is exploited by the HARQ in the fast-fading (resp. slow-fading) condition compared to the slow-fading (resp. quasi-static) conditions and, consequently, different outage probability constraints are satisfied with fewer antennas in the fast-fading (resp. slow-fading) conditions (Fig. 2). However, the gap between the system performance in different fading conditions decreases with the number of antennas (Fig. 2).
  \item In Cases 2 with many antennas only at the transmitter, the performance improvement becomes limited when the number of transmit antennas reaches, say, $N_\text{t}\gtrsim 70$. In the meantime, considerable improvement is achieved by adding more antennas at the receiver (Fig. 2. Also, similar arguments hold for Case 1 although not demonstrated in the figures).
  \item The HARQ reduces the required number of antennas significantly (Fig. 3). For instance, consider the quasi-static conditions, the outage probability constraint $\Pr(\text{Outage})\le 10^{-4},$ $N_\text{t}=5,$ $\phi=5\text{ dB}$ and the code rate $20$ npcu. Then, the implementation of HARQ with a maximum of $M=2$ retransmissions reduces the required number of receive antennas from $95$ without HARQ to $15$ (Fig. 3). Moreover, the effect of HARQ increases with the number of transmit/receive antennas (Fig. 3).
  \item Different outage probability requirements are satisfied with relatively few antennas. For instance, consider a SIMO setup in quasi-static conditions and $M=1, \phi=5\text{dB}.$ Then, with an initial rate $R=3$ npcu, the outage probabilities $\Pr(\text{Outage})\le 10^{-3}, 10^{-4}$ and $10^{-5}$ are satisfied with $16, 18$ and $20$ receive antennas, respectively (Fig. 5a). These numbers increase to $31, 35$, and $38$ for $R=4$ npcu (Fig. 5a).
  \item The normalized outage factor, i.e., the negative of the slope of the outage probability curve versus the product of the number of antennas as the number of antennas increases,  follows the theoretical results of Theorem 2 with high accuracy (Figs. 5b and 5c). Moreover, the number of fading realizations experienced during the packet transmission increases the normalized outage factor linearly (Fig. 5c. Also, see Theorem 2 and its following discussions).
  \item The inefficiency of the PAs affects the performance of large MIMO setups remarkably. For instance, with the parameter settings of Fig. 6a and $R=10\text{ npcu}, N_\text{r}=2,$ the inefficiency of the PAs increases the consumed power by $\sim 11\text{ dB}$ (Fig. 6a). However, the effect of the PAs inefficiency decreases with the SNR which is intuitively because the \emph{effective} efficiency of the PAs $\epsilon^\text{effective}=\epsilon(\frac{\phi}{\phi^\text{max}})^\vartheta$ is improved at high SNRs. On the other hand, adaptive power allocation between the HARQ retransmissions reduces the required number of antennas marginally (Fig. 6b). Therefore, considering Fig. 6b and the implementation complexity of adaptive power allocation, non-adaptive power allocation is a good choice for large MIMO systems.
\end{itemize}

\subsection{On the Effect of Spatial Correlation}
Throughout the paper, we considered IID fading conditions motivated by the fact that the millimeter-wave communication, which will definitely be a part in the next generation of wireless networks, makes it possible to assemble many antennas close together with negligible spatial correlations \cite{6515173,5783993}. However, it is still interesting to analyze the effect of the antennas spatial correlation on the system performance. For this reason, considering Case 2 with $N_\text{r}=1$, Fig. 7 demonstrates the required number of antennas in the spatially-correlated conditions where, denoting the transpose operator by $()^\text{T},$ the successive elements of the channel vector $\textbf{H}=[h_1,\ldots,h_{N_\text{t}}]^\text{T}$ follow
\begin{align}
h_i=\beta h_{i-1}+\sqrt{1-\beta^2}\varpi, \varpi\sim\mathcal{CN}(0,1), h_0\sim\mathcal{CN}(0,1).
\end{align}
Here, $\beta$ is a correlation coefficient where $\beta=0$ (resp. $\beta=1$) corresponds to the uncorrelated (resp. fully correlated) conditions. This is a well-established
model considered in the literature for different applications, e.g., \cite{4801449}.

As shown in the figure, the effect of the antennas spatial correlation on the required number of antennas is negligible for correlation coefficients of, say, $\beta\lesssim 0.4$.  This is in harmony with, e.g., \cite{6612899,1673668} which, with different problem formulations/metrics, derive the same conclusion about the effect of the antennas correlation on the system performance. Then, the sensitivity to the spatial correlation increases for large values of the correlation coefficients, and the required number of antennas increases with $\beta$. However, the important point is that the curves follow the same trend, for a large range of correlation coefficients (Fig. 7). Thus, with high accuracy, the same scaling laws as in the IID scenario also hold for the correlated conditions, as long as the correlation coefficient is not impractically high. Moreover, we observe the same conclusions in the other cases, although not demonstrated in the figure. Finally, it is worth noting that, as shown in \cite{1237135}, for moderate/large number of transmit and/or receive antennas and with appropriate mean and variance selection, the accumulated mutual information of the correlated MIMO setups follows  Gaussian distributions with high accuracy. Therefore, one can use \cite{1237135} and the same procedure as in our paper to derive closed-form expressions for the required number of antennas in the spatially-correlated MIMO-HARQ systems.

\section{Conclusion}
This paper studied the required number of antennas satisfying different outage probability constraints in large but finite MIMO setups. We showed that different quality-of-service requirements can be satisfied with relatively few transmit/receiver antennas. Also, we derived closed-form expressions for the normalized outage factor which is defined as the negative of the slope of the outage probability curve plotted versus the product of number of antennas. As demonstrated, the required number of antennas decreases by the implementation of HARQ remarkably. The effect of the antennas spatial correlation on the required number of antennas is negligible for small/moderate correlation coefficients, while its effect increases in highly correlated conditions. Finally, with the problem formulation of the paper, the performance of the large MIMO systems is sensitive (resp. almost insensitive) to the power amplifiers inefficiency (resp. adaptive power allocation between the HARQ retransmissions). Performance analysis in the cases with other HARQ protocols is possibly an interesting extension of the work presented in this paper.

\begin{figure}\label{fig:NPA}
\vspace{-3mm}
\centering
  \includegraphics[width=.87\columnwidth]{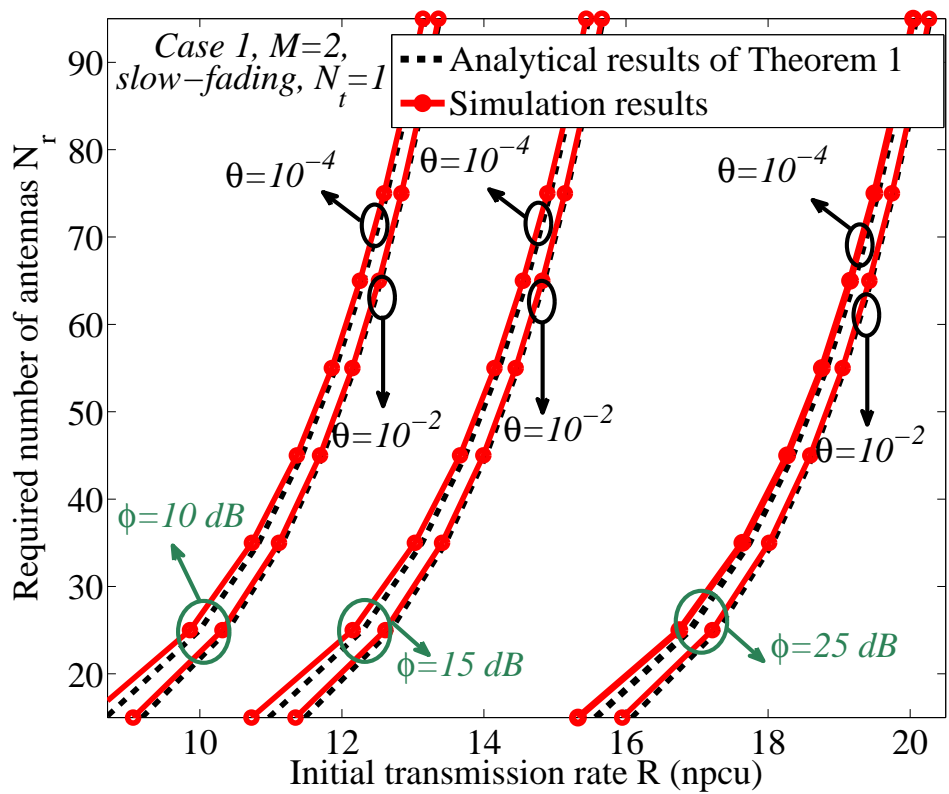}\\\vspace{-2mm}
\caption{The required number of receive antennas vs the initial transmission rate $R$ (Case 1: large $N_\text{r}$, given $N_\text{t}$). Outage probability constraint $\Pr(\text{Outage})<\theta\,(\theta=10^{-4}$ or $10^{-2})$, slow-fading conditions, $M=2,$ and $N_\text{t}=1.$ }\label{figure111}
\vspace{-3mm}
\end{figure}

\begin{figure}\label{fig:NPA}
\vspace{-3mm}
\centering
  \includegraphics[width=.87\columnwidth]{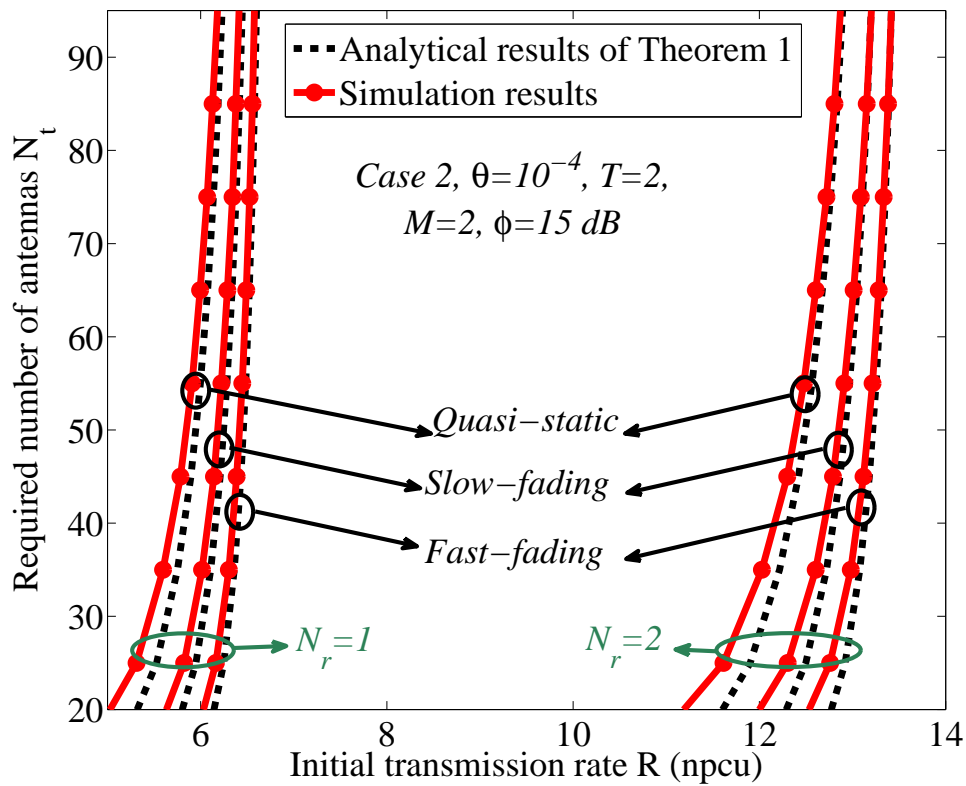}\\\vspace{-2mm}
\caption{The required number of transmit antennas vs the initial transmission rate $R$ for the quasi-static, slow- and fast-fading conditions (Case 2: large $N_\text{t}$, given $N_\text{r}$). Outage probability constraint $\Pr(\text{Outage})<\theta,\,(\theta=10^{-4}$), $\phi=15 \text{ dB, }T=2, M=2,$ and $N_\text{r}=1$ or $2.$ }\label{figure111}
\vspace{-3mm}
\end{figure}

\begin{figure}\label{fig:NPA}
\vspace{-3mm}
\centering
  \includegraphics[width=.87\columnwidth]{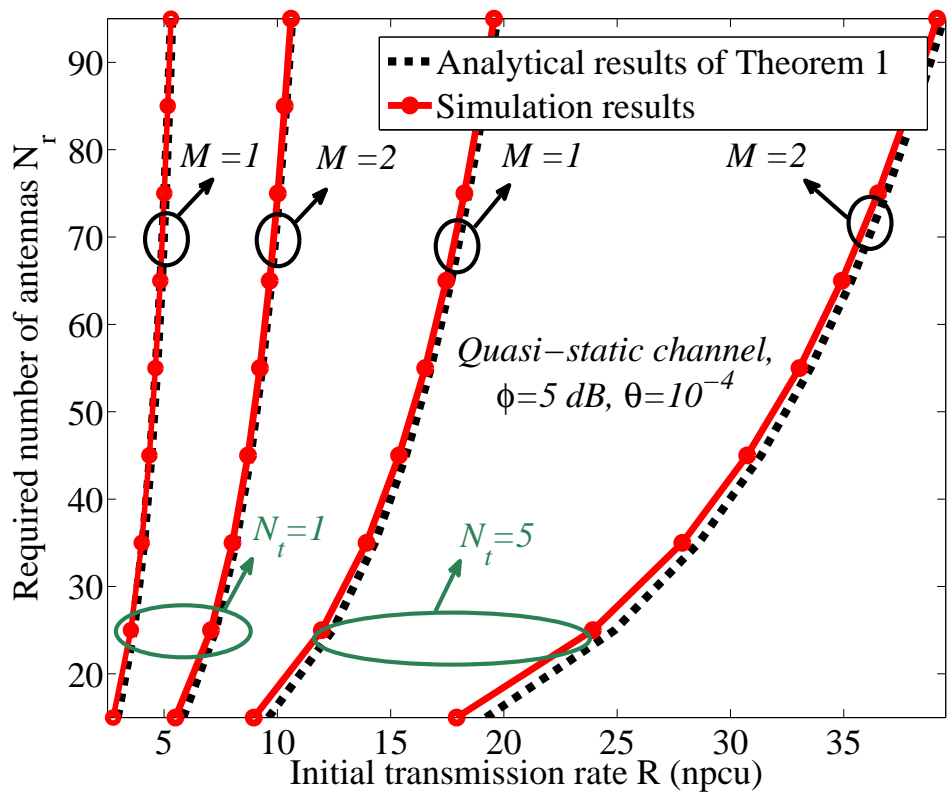}\\\vspace{-2mm}
\caption{The required number of transmit antennas in the scenarios with HARQ ($M=2$) and without HARQ ($M=1$), Case 1: (large $N_\text{r}$, given $N_\text{t}$). Outage probability constraint $\Pr(\text{Outage})<\theta$ with $\theta=10^{-4}$, $\phi=5 \text{ dB, }, N_\text{t}=1$ or $5,$ and quasi-static conditions.}\label{figure111}
\vspace{-3mm}
\end{figure}

\begin{figure}\label{fig:NPA}
\vspace{-3mm}
\centering
  \includegraphics[width=.87\columnwidth]{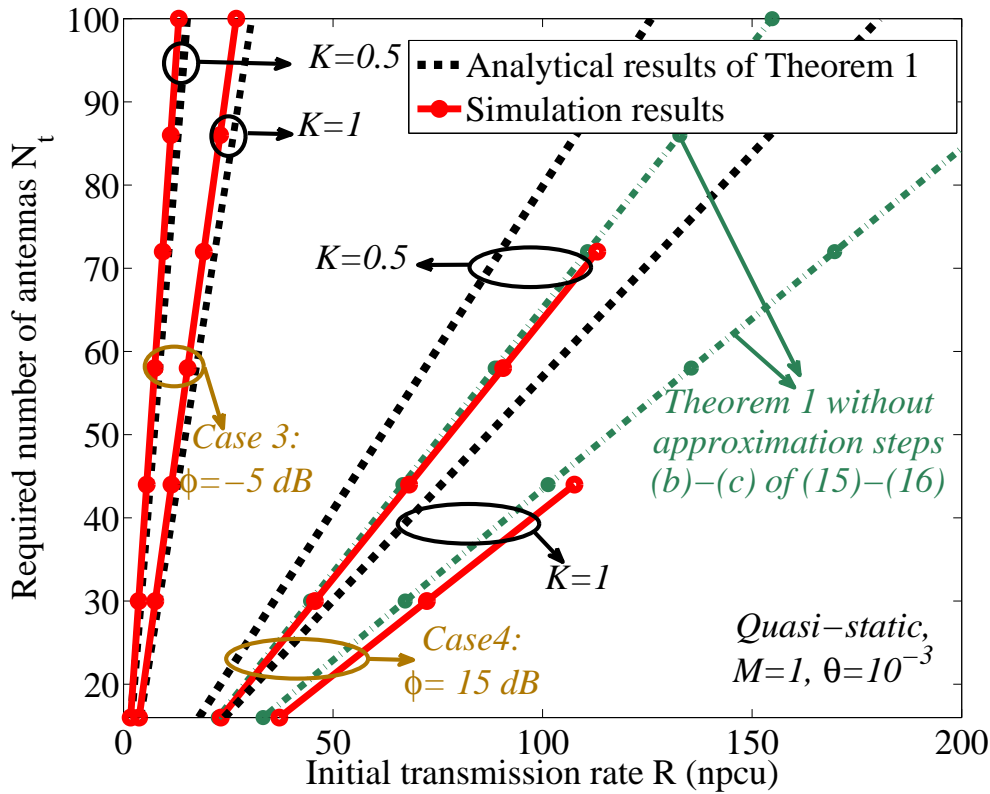}\\\vspace{-2mm}
\caption{The required number of transmit antennas vs the initial transmission rate, Cases 3 and 4: (large $N_\text{t}, N_\text{r},\frac{N_\text{t}}{N_\text{r}}=K$). Outage probability constraint $\Pr(\text{Outage})<\theta$ with $\theta=10^{-3}$, $\phi=-5$ or $15 \text{ dB, }, M=1,$ and quasi-static conditions.}\label{figure111}
\vspace{-3mm}
\end{figure}

\begin{figure}\label{fig:NPA}
\vspace{-3mm}
\centering
  \includegraphics[width=.99\columnwidth]{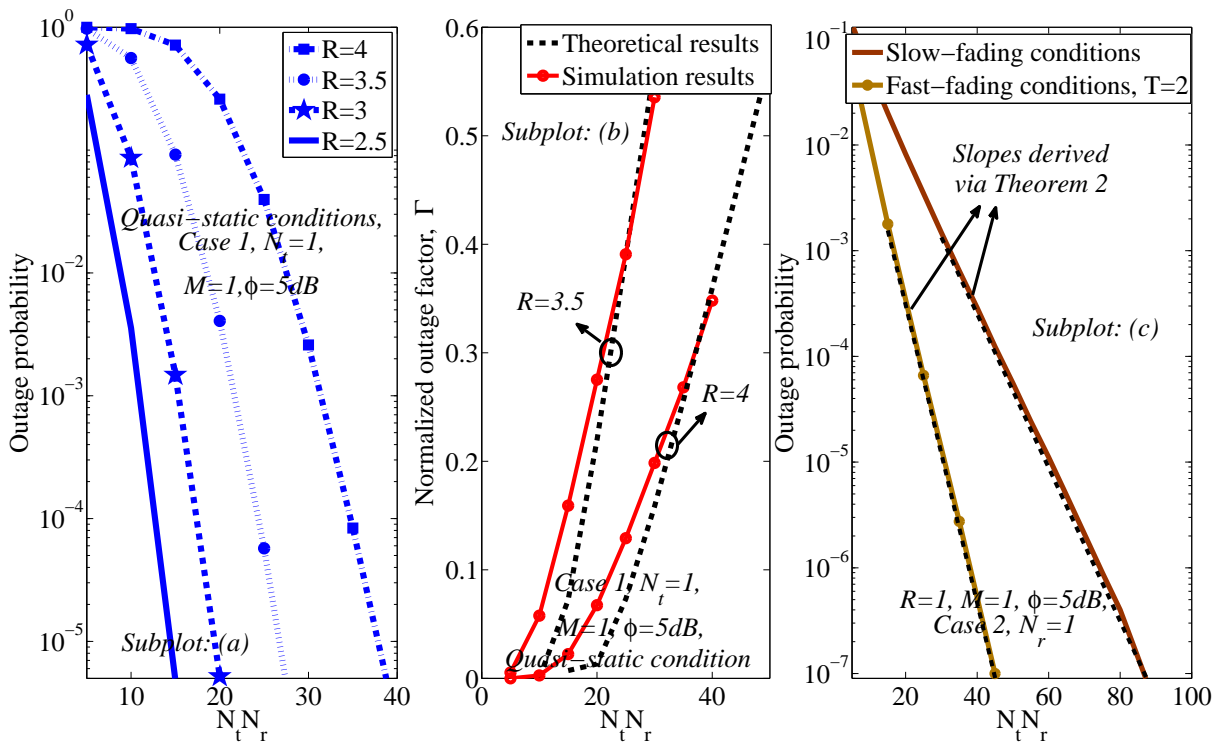}\\\vspace{-2mm}
\caption{Subplot (a): The outage probability vs the product of the number of transmit and receive antennas, Case 1, $N_\text{t}=1,$ quasi-static conditions, $M=1, \phi=5\text{ dB}.$ Subplot (b): The normalized outage factor vs the product of the number of transmit and receive antennas, Case 1, $N_\text{t}=1,$ quasi-static conditions, $M=1,$ and $ \phi=5\text{ dB}.$ Subplot (c): The outage probability vs the product of the number of transmit and receive antennas, Case 2, $N_\text{r}=1,$ $M=1,$ $\phi=5\text{ dB},$ and $R=1.$}\label{figure111}
\vspace{-3mm}
\end{figure}

\begin{figure}\label{fig:NPA}
\vspace{-3mm}
\centering
  \includegraphics[width=.87\columnwidth]{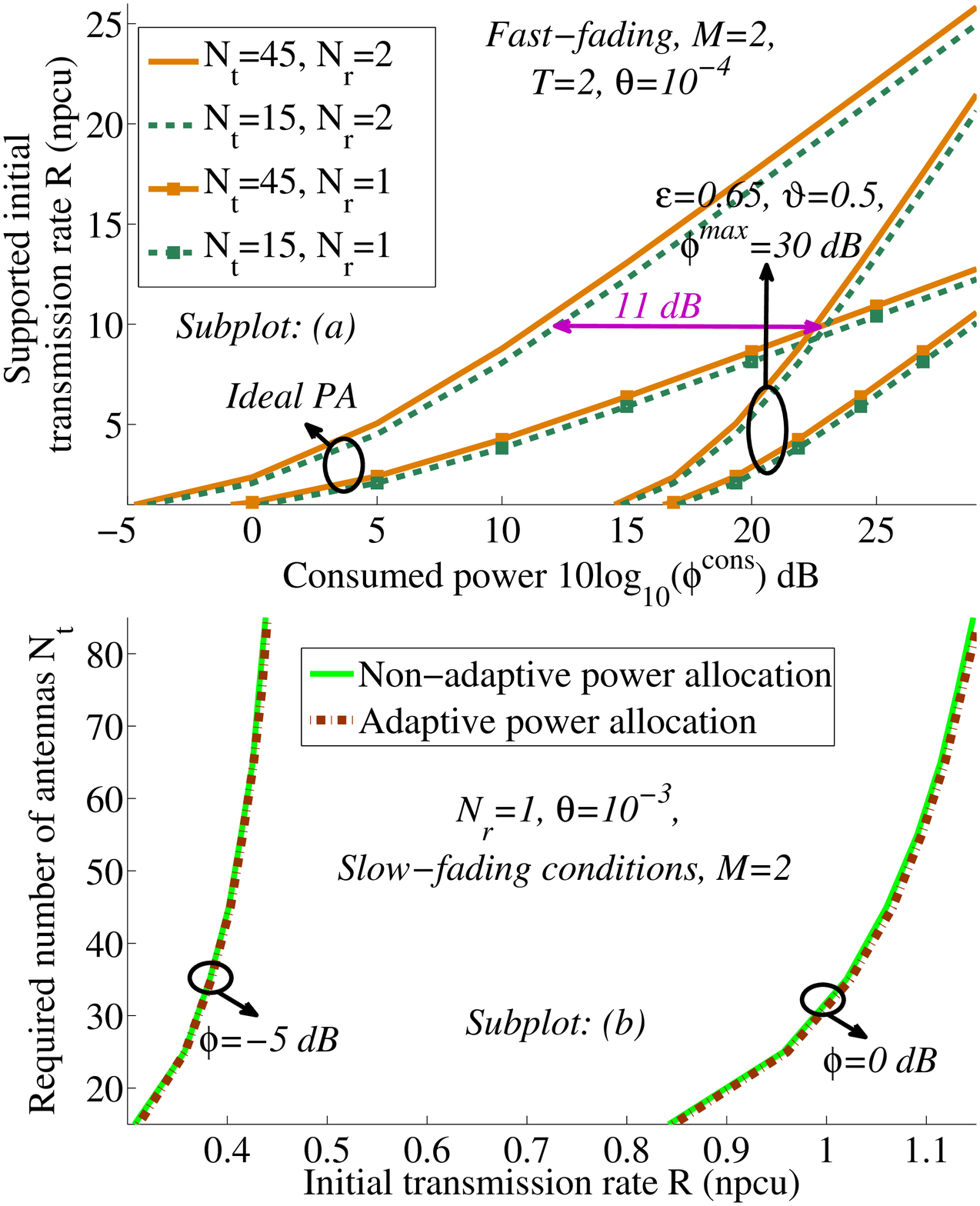}\\\vspace{-2mm}
\caption{Subplot (a): On the effect of non-ideal PAs. Supported initial transmission rate vs the consumed power $10\log_{10}(\phi^{\text{cons}})$. Case 2: (large $N_\text{t}$, given $N_\text{r}$), outage probability constraint $\Pr(\text{Outage})<\theta,\,\theta=10^{-4}$, $T=2, M=2,$ fast-fading conditions. In the cases with non-ideal PAs, we set $\epsilon=0.65, \vartheta=0.5, \phi^{\text{max}}=30\text{ dB.}$ Subplot (b): On the effect of adaptive power allocation. The required number of transmit antennas vs the initial transmission rate $R$ (npcu). Case 2: (large $N_\text{t}$, given $N_\text{r}$), outage probability constraint $\Pr(\text{Outage})<\theta$ with $\theta=10^{-3}$, $N_\text{r}=1, \phi=-5$ or $0 \text{ dB, } M=2,$ and slow-fading conditions.}\label{figure111}
\vspace{-3mm}
\end{figure}

\begin{figure}\label{fig:NPA}
\vspace{-3mm}
\centering
  \includegraphics[width=.87\columnwidth]{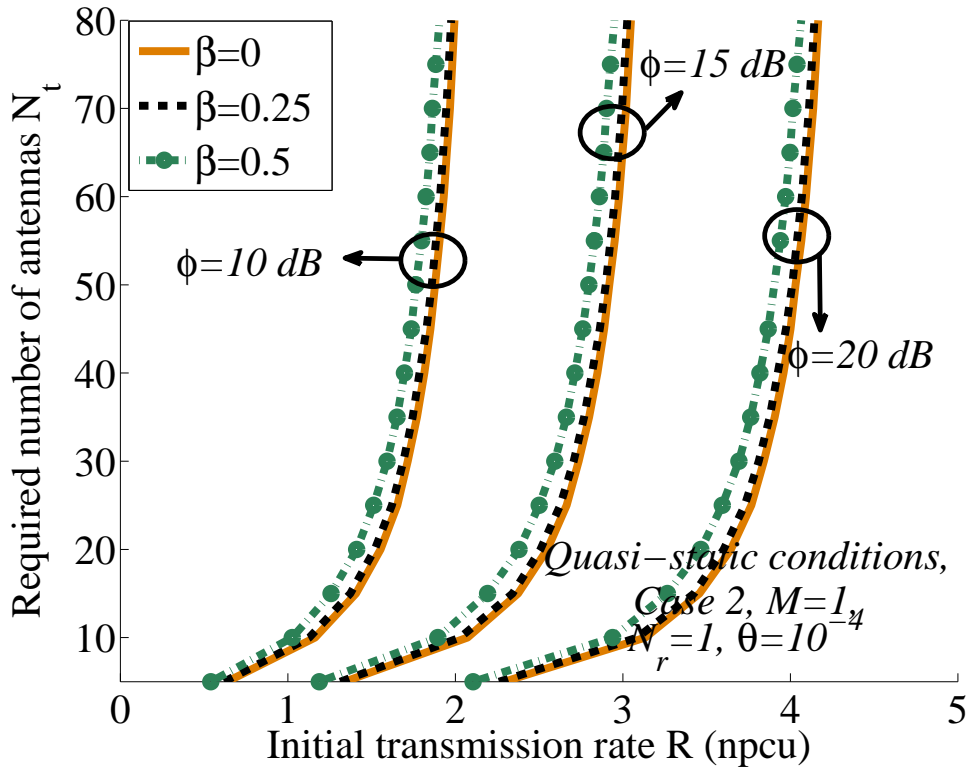}\\\vspace{-2mm}
\caption{The required number of antennas in different spatially-correlated conditions. Case 2: (large $N_\text{t}$, given $N_\text{r}$), outage probability constraint $\Pr(\text{Outage})<\theta$ with $\theta=10^{-4}$, $M=1,$ quasi-static conditions, and $N_\text{r}=1$. }\label{figure111}
\vspace{-3mm}
\end{figure}
\vspace{-4mm}
\bibliographystyle{IEEEtran} 
\bibliography{masterlargeMIMO}

\vfill

\end{document}